\documentclass[12pt]{article}
\usepackage{latexsym,amsmath}
\begin{document}
\begin{center}
{\bf \Large Crossing of the w\,=\,-1 barrier in two-fluid viscous
modified
gravity} \\

\bigskip

\bigskip

Iver Brevik\footnote{Department of Energy and Process Engineering,
Norwegian University of Science and Technology, N-7491 Trondheim,
Norway. Email: iver.h.brevik@ntnu.no}
\end{center}

\bigskip

\begin{abstract}

Singularities in the dark energy late universe are discussed,
under the assumption that the Lagrangian contains the Einstein
term $R$ plus a modified gravity term of the form $R^\alpha$,
where $\alpha$ is a constant. It is found, similarly as in the
case of pure Einstein gravity [I. Brevik and O. Gorbunova, Gen.
Rel. Grav. {\bf 37}, 2039 (2005)], that the fluid can pass from
the quintessence region ($w>-1$) into the phantom region ($w<-1$)
as a consequence of a bulk viscosity varying with time. It becomes
necessary now, however, to allow for a two-fluid model, since the
viscosities for the two components vary differently with time. No
scalar fields are needed for the description of the passage
through the phantom barrier.

\end{abstract}

\bigskip


\bigskip
\begin{center}
\today
\end{center}

\section{Introduction}

The possibility of crossing the $w=-1$ barrier in dark energy
physics is a topic that has attracted a great deal of interest. It
is usually assumed that the equation of state for the cosmic fluid
can be written in the form
\begin{equation}
p=w\rho \equiv (\gamma-1)\rho. \label{1}
\end{equation}
Here the equation of state parameter $w=-1$ or $p=-\rho$
corresponds to a "vacuum fluid", with bizarre thermodynamic
properties such as possibly negative entropies \cite{brevik04}.
Cosmological observations such as SNe Ia \cite{riess04}, WMAP
\cite{bennett03}, SDSS \cite{tegmark04} and the Chandra X-ray
Observatory \cite{allen04} indicate that the present universe is
accelerating. Moreover, based upon the observed data it has been
conjectured that $w$ is a varying function of time. For instance,
as discussed in Ref.~\cite{vikman05}, $w$ might have been around 0
at redshift $z \sim 1$ and may be around -1.2 today. (Cf. also the
very recent analysis of observational constraints on the dark
energy \cite{capozziello05}.) Perhaps is $w$ even an oscillating
function in time. In view of these circumstances the analysis of a
possible crossing of the phantom barrier $w=-1$, from the
quintessence region $(-1<w<-1/3)$ into the phantom region $w<-1$,
is obviously of physical interest. It ought to be noted that both
quintessence and phantom fluids lead to the inequality $\rho+3p
\leq 0$, thus breaking the strong energy condition.

The occurrence of a phantom fluid leads inevitably to a
singularity in the future, called the Big Rip
\cite{caldwell03,mcinnes02,barrow04,nojiri05}. Often, the crossing
of the phantom barrier is described in terms of one or more scalar
fields, besides the gravitational field. For instance, as shown
recently in Ref.~\cite{capozziello06}, one can consider inflation,
dark energy and dark matter under the same standard:
phantom-nonphantom transitions may appear such that the universe
could have been effectively quantum-like both at early times and
will behave similarly at late times. Cf. also the related papers
\cite{capozziello05,li05}. Here, only one scalar field is
considered. In several papers - cf., for instance,
Refs.~\cite{nojiri05a,wei06,feng05,zhao05,elizalde05} -
cosmological models with two scalar fields (or fluids) are
considered. In Ref.~ \cite{vikman05}, arguments based upon a
stability analysis are given why phantom barrier crossing cannot
be effectuated in terms of one scalar field alone.

One interesting alternative to a scalar field theory is to allow
for some sort of modification of the standard Einstein theory. One
kind of approach is to introduce higher order derivatives of
scalar fields in the Lagrangian \cite{li05,anisimov05}. The form
of modified gravity that we shall be concerned with here is to
 introduce a term $R^\alpha$ in the
gravitational action, where $\alpha$ is a constant.  A recent
introduction to modified gravity of this kind (including also
Gauss-Bonnet gravity) is given by Nojiri and Odintsov
\cite{nojiri06}. Some other papers dealing with modified gravity
from different viewpoints are
Refs.~\cite{gognola06,nojiri03,carroll04,carter05,barrow87,clifton05}.

The body of literature in this field of research is large, and the
above list of references is not intended to be exhaustive. We turn
now, however, to the main theme of the present paper, which is to
introduce a {\it bulk viscosity} $\zeta$ in the cosmic fluid. Bulk
viscosity is compatible with a universe that is spatially
isotropic. We shall take  $\zeta$ to be dependent on time. In the
paper \cite{brevik05} dealing with Einstein's gravity, we showed
how the ansatz of letting $\zeta$ be proportional to the scalar
expansion $\theta =3H$ can drive the fluid into the phantom region
even if it starts from the quintessence region in the non-viscous
case. In Ref.~\cite{brevik05a} we considered a more general
situation with the action
\begin{equation}
S=\frac{1}{2\kappa^2} \int d^4 x \sqrt{-g}\,(f_0R^\alpha +L_m),
\label{2}
\end{equation}
where $\kappa^2=8\pi G$, $f_0$ and $\alpha$ are constants, and
$L_m$ is the matter Lagrangian. This is the kind of action
recently analyzed by Abdalla {\it et al.} \cite{abdalla05}. Based
upon a natural assumption for the time variation of the scale
factor, we verified that the field equations are satisfied for
general $\alpha$, and we investigated the ansatz of letting
$\zeta$ be proportional to $\theta^{2\alpha -1}$. This kind of
modified gravity was further considered in Ref.~\cite{brevik05b}.
We found the same kind of behavior also in this case: the fluid
can in principle  make a viscosity-generated passage from the
quintessence region into the phantom region. There are some
obvious advantages of this kind of theory: the theory is
mathematically simple, it makes use of standard concepts from
fluid mechanics only (in addition to gravity), and there is no
need of introducing a scalar field.

The approach of Ref.~\cite{brevik05b} is however physically
incomplete, in the following sense: it involves the gravitational
correction term $R^\alpha$ only, in the Lagrangian.  It would be
more appropriate to include the Einstein term $R$ also, so that
the action takes the form
\begin{equation}
S=\frac{1}{2\kappa^2}\int d^4 x \sqrt{-g}\,(R+f_0R^\alpha +L_m).
\label{3}
\end{equation}
This form of action is the starting point for the investigation in
the present paper. As we shall see, the fluid possesses
essentially   the same property as before; it can slide through
the phantom barrier into the phantom region as a consequence of
viscosity. But there is one essential difference: the theory now
requires a {\it two-component} model of the fluid. The two
components belonging to the Einstein gravity and to the modified
gravity vary differently with time. It becomes natural to
speculate if this necessity of dividing the cosmic fluid into two
components is not after all another demonstration of the
difficulties with a one-component model
 that have been found previously in other
investigations, within a conventional scalar field approach.

The present series of works is of course not the first time that
the viscosity concept has been introduced in cosmology. Misner was
probably  the first to introduce viscosity in the cosmic fluid in
his study of how initial anisotropies in the early universe become
relaxed \cite{misner68}. Another early paper is that of
Padmanabhan and Chitre \cite{padmanabhan87}. An extensive review
of the development up to 1990 is given by Gr{\o}n \cite{gron90}.
We have ourselves dealt earlier with viscous entropy production in
the early universe \cite{brevik94}, and with viscous fluids on the
Randall-Sundrum branes \cite{brevik04c,brevik05c}. Ren and Meng
recently discussed a cosmological model with a dark viscous fluid
described by an effective equation of state, and compared with SNe
data \cite{ren06}. Cataldo {\it et al.} considered viscous dark
energy and phantom evolution within the framework of the Eckart
theory of relativistic irreversible thermodynamics
\cite{cataldo05}. Kofinas {\it et al.} considered the crossing of
the $w=-1$ barrier from a brane-bulk energy exchange scenario
containing an induced gravity curvature correction term
\cite{kofinas05}. As discussed by Nojiri and Odintsov
\cite{nojiri05a} and by Capozziello {\it et al.}
\cite{capozziello05}, a dark fluid with a time dependent bulk
viscosity can be considered as a fluid with an inhomogeneous
equation of state.

In the next section we present the field equations, and the
conservation equation for energy, and derive herefrom two
expressions, Eqs.~(\ref{16}) and (\ref{17}), for the coefficient
$B$ defined in Eq.~(\ref{12}). A positive value of $B$ leads to a
Big Rip. Consistency of the formalism requires that the
coefficients $\tau_E$ and $\tau_\alpha$ introduced in
Eqs.~(\ref{14}) and (\ref{15}) are related to each other. Examples
with positive values of $\alpha$ are discussed in Sect.~3. The
case of negative $\alpha$, discussed in Sect.~4, is exceptional,
since the physical quantities associated with the modified fluid
component does not go to infinity at Big Rip, but rather to zero.
Still, the Hubble factor, as well as the physical quantities
associated with the Einstein component, diverge.

\section{General formalism}

We assume the spatially flat FRW metric
\begin{equation}
ds^2=-dt^2+a^2(t)d{\bf x}^2, \label{4}
\end{equation}
put the cosmological constant $\Lambda$ equal to zero, and adopt
the energy-momentum tensor of the viscous fluid in the standard
form
\begin{equation}
T_{\mu\nu}=\rho U_\mu U_\nu +\tilde{p}h_{\mu\nu}, \label{5}
\end{equation}
where $h_{\mu\nu}=g_{\mu\nu}+U_\mu U_\nu$ is the projection tensor
and $\tilde{p}=p-\zeta \theta$ the effective pressure.  In
comoving coordinates, the four-velocity of the fluid is $U^0=1,\,
U^i=0$. From variation of the action (\ref{3}) we obtain the
equations of motion
\[ -\frac{1}{2}g_{\mu\nu}R(1+f_0R^{\alpha-1})+R_{\mu\nu}(1+\alpha
f_0R^{\alpha-1}) \]
\begin{equation}
-\alpha f_0\nabla_\mu\nabla_\nu R^{\alpha-1}+\alpha f_0
\,g_{\mu\nu}\nabla^2 R^{\alpha-1}=\kappa^2T_{\mu\nu}, \label{6}
\end{equation}
where $T_{\mu\nu}$ corresponds to the term $L_m$ in the
Lagrangian.

Of main interest is the (00)- component of this equation. Using
that
\begin{equation}
R_{00}=-3\ddot{a}/a,\quad R=6(\dot{H}+2H^2), \label{7}
\end{equation}
as well as $T_{00}=\rho$, we obtain
\[ 3H^2+\frac{1}{2}f_0 R^\alpha-3\alpha f_0(\dot{H}+H^2)R^{\alpha-1}
\]
\begin{equation}
+3\alpha (\alpha-1)f_0HR^{\alpha-2}\dot{R}=\kappa^2\rho. \label{8}
\end{equation}
An important property of Eq.~(\ref{6}) is that the covariant
divergence of the LHS is equal to zero \cite{koivisto05}. Thus
\begin{equation}
\nabla^\nu T_{\mu\nu}=0, \label{9}
\end{equation}
just as in the case of Einstein's gravity. Energy-momentum
conservation is a consequence of the field equations. Contracting
the expression (\ref{5}) for the energy-momentum tensor with
$U^\mu$ we obtain the energy conservation equation
\begin{equation}
\dot{\rho}+(\rho+p)3H=9\zeta H^2. \label{10}
\end{equation}
We now differentiate the expression (\ref{8}) with respect to $t$,
and insert $\dot{\rho}$ from Eq.~(\ref{10}). After some
calculation we then obtain \[ 6\dot{H}+9\gamma
H^2+\frac{3}{2}\gamma f_0R^\alpha -3\alpha
f_0[(3\gamma-2)\dot{H}+3\gamma H^2]R^{\alpha-1} \]
\begin{equation}
+3\alpha
(\alpha-1)f_0[(3\gamma-1)H\dot{R}+\ddot{R}]R^{\alpha-2}+3\alpha(\alpha-1)(\alpha-2)f_0\dot{R}^2R^{\alpha-3}=9\kappa^2\zeta
H. \label{11}
\end{equation}
Recalling that $R=6(\dot{H}+2H^2)$, we see that this is a
nonlinear differential equation for $H(t)$. We shall seek for
solutions that are compatible with the following basic ansatz for
the Hubble parameter:
\begin{equation}
H=\frac{H_0}{X}, \quad {\rm where} \quad X \equiv 1-BH_0 t.
\label{12}
\end{equation}
Here $B$ is a nondimensional quantity whose value is dependent on
$\alpha$. We take the initial time to be $t_0=0$, and give a
subscript zero to quantities referring to this instant. If Big Rip
shall occur ($H\rightarrow \infty$), $B$ has to be positive.

From the mathematical structure of Eq.¨(\ref{11}) it is clear that
we cannot eliminate the time dependent terms in the governing
equation for $B$ simply by letting $\zeta$ be proportional to a
power of the scalar expansion $\theta$. This feature contrasts
that found in earlier papers \cite{brevik05} and \cite{brevik05b}.
It becomes now necessary to allow for a {\it two-fluid} model. We
shall write $\zeta =\zeta(t)$ as a sum of a term $\zeta_E$
referring to Einstein gravity and a term $\zeta_\alpha$ referring
to modified gravity:
\begin{equation}
\zeta=\zeta_E+\zeta_\alpha, \label{13}
\end{equation}
where
\begin{equation}
\zeta_E=\tau_E\theta =3\tau_EH, \label{14}
\end{equation}
\begin{equation}
\zeta_\alpha=\tau_\alpha \theta^{2\alpha-1}=\tau_\alpha
(3H)^{2\alpha-1}, \label{15}
\end{equation}
$\tau_E$ and $\tau_\alpha$ being constants. Then Eq.~(\ref{11})
can be satisfied for the Einstein part and the modified part
separately. We get in this way two different algebraic expressions
determining the constant $B$:
\begin{equation}
B=-\frac{3}{2}\gamma+\frac{9}{2}\kappa^2\tau_E, \label{16}
\end{equation}
\[
(B+2)^{\alpha-1}
\Big\{9(2-\alpha)\gamma+3[\alpha+3\gamma+\alpha(2\alpha-3)(3\gamma-1)]B
\]
\begin{equation}
+6\alpha(\alpha-1)(2\alpha-1)B^2\Big\}=\frac{18\kappa^2}{f_0}\left(\frac{3}{2}\right)^\alpha
\tau_\alpha; \label{17} \end{equation}
 the time-dependent terms
drop out. If $f_0=0$ (Einstein gravity only), Eq.~(\ref{16}) is in
accordance with Ref.~\cite{brevik05}. If the Einstein term is
absent, Eq.~(\ref{17}) agrees with Ref.~\cite{brevik05b}.

One important property follows at once from the compatibility of
Eqs.~(\ref{16}) and (\ref{17}): there must be a relationship
between $\tau_E$ and $\tau_\alpha$. This relationship $\tau_\alpha
=\tau_\alpha(\tau_E)$ depends on the values of $\alpha$ and $f_0$,
as well as on the fluid parameter $\gamma=w+1$. There seems to be
no direct physical reason behind this dependence; it results
simply from the mathematical consistency of the model.

In the next section we shall turn to considering specific models.
We note, however, the general expressions for $B$ that follow
directly from the energy conservation equation (\ref{10}): as this
equation has to be fulfilled for the two fluid components
separately,
\begin{equation}
\dot{\rho}_E+(\rho_E+p_E)3H=9\zeta_E H^2, \label{18}
\end{equation}
\begin{equation}
\dot{\rho}_\alpha+(\rho_\alpha+p_\alpha)3H=9\zeta_\alpha H^2,
\label{19}
\end{equation}
we get
\begin{equation}
B=-\frac{3\gamma}{2}+\frac{27\tau_E}{2}\frac{H_0^2}{\rho_{0E}},
\label{20}
\end{equation}
\begin{equation}
B=-\frac{3\gamma}{2\alpha}+\frac{3\tau_\alpha}{2\alpha}\frac{(3H_0)^{2\alpha}}{\rho_{0\alpha}}.
\label{21}
\end{equation}
We here made use of the time dependent relations
\begin{equation}
\zeta_\alpha=\tau_\alpha \left(\frac{3H_0}{X}\right)^{2\alpha-1},
\quad \rho_E=\frac{\rho_{0E}}{X^2}, \quad
\rho_\alpha=\frac{\rho_{0\alpha}}{X^{2\alpha}}, \label{22}
\end{equation}
In order to calculate $B$ from the expressions (\ref{20}) or
(\ref{21}), at least one of the initial densities $\rho_{0E}$ or
$\rho_{0\alpha}$ have to be known.

Equation (\ref{20}) agrees with Eq.~(\ref{16}) in view of the
first Friedmann equation $3H_0^2= \kappa^2\rho_{0E}$ at $t=0$.

\section{Positive values of $\alpha$: Some examples}

\subsection{Small deviations from Einstein's gravity}

It is natural, for exemplification, to start with a modified
gravity system where the exponent $\alpha$ in the Lagrangian
(\ref{3}) is small. We shall take $\alpha$ to be positive, and
also assume that  $f_0$ is positive and small:
\begin{equation}
\alpha \ll 1,\quad f_0 \ll 1. \label{23}
\end{equation}
As approximatively  $R^\alpha =1+\alpha \ln R$, we see that
 terms containing $f_0 \alpha$ are of second order and thus
 negligible.
The product $f_0R^\alpha$ can thus simply be replaced with $f_0R$,
and the action (\ref{3})  reduces to
\begin{equation}
S=\frac{1}{2\kappa^2}\int d^4 x \sqrt{-g}[(1+f_0)R+L_m].
\label{24}
\end{equation}
From this equation, or directly from Eq.~(\ref{6}), it follows
that the Ricci tensor can be taken to be the same as in Einstein's
gravity. The energy-momentum tensor becomes however rescaled with
the factor $1/(1+f_0)$ (or $1-f_0$). The field equations (\ref{6})
reduce to
\begin{equation}
-\frac{1}{2}g_{\mu\nu}R+R_{\mu\nu}=\frac{\kappa^2
T_{\mu\nu}}{1+f_0}. \label{25}
\end{equation}
The first Friedmann equation (corresponding to $\mu=\nu=0$) now
becomes
\begin{equation}
3H^2=\frac{\kappa^2}{1+f_0}\rho. \label{26}
\end{equation}
As for the determination of the constant $B$ we obtain for the
Einstein component the same equation (\ref{16}) as before. For the
$\alpha$-component (\ref{17}), when combined with (\ref{16}), we
obtain the simple relation
\begin{equation}
\tau_\alpha=f_0\tau_E, \label{27}
\end{equation}
valid for all values of $\gamma$.  Note the dimensions:
$[f_0]={\rm cm}^{2(\alpha-1)},\, [\tau_E]={\rm
cm}^{-2},\,[\tau_\alpha]={\rm cm}^{2(\alpha-2)}$. (For clarity we
write $\tau_\alpha$ instead of $\tau_1$.) Under the present
conditions, $\tau_\alpha \ll \tau_E$.

Already from Eq.~(\ref{16}) it is clear that $B$ can be greater
than zero, thus leading to a Big Rip, if $\tau_E$ is big enough.
Let us summarize the time dependencies of the physical quantities:
\begin{equation}
\zeta_E=\frac{3\tau_EH_0}{X},\quad \rho_E=\frac{\rho_{0E}}{X^2},
\label{28}
\end{equation}
\begin{equation}
\zeta_\alpha=\frac{3\tau_Ef_0H_0}{X}, \quad
\rho_\alpha=\frac{\rho_{0\alpha}}{X^2}, \label{29}
\end{equation}
with $X=1-BH_0 t$ as before. Moreover, $\rho=\rho_E+\rho_\alpha$
for all $t$.

\subsection{The case $\alpha =1/2$}
This case, corresponding to
\begin{equation}
S=\frac{1}{2\kappa^2}\int d^4 x \sqrt{-g}\left( R+f_0\sqrt{R}+L_m
\right), \label{30}
\end{equation}
is  mathematically simplifying, and can moreover be of physical
interest. Equation (\ref{17}) yields the following  quadratic
equation for $B$:
\begin{equation}
B^2+6(\gamma-4T^2)B+9\gamma^2-48 T^2=0, \label{31}
\end{equation}
with
\begin{equation}
T=\frac{\kappa^2}{f_0}\tau_\alpha. \label{32}
\end{equation}
The product of the two roots for $B$ is equal to
$9\gamma^2-48T^2$. If we put $\gamma=0$, which is the most
interesting case, we thus see that the product is always negative
and there is one positive and one negative root.  The positive
root necessarily leads to a Big Rip. When $\gamma=0$ we get from
Eqs.~(\ref{31}) and (\ref{32}), when taking Eq.~(\ref{16}) into
account,
\begin{equation}
\tau_\alpha=\frac{9f_0\tau_E}{4\sqrt{12+27\kappa^2 \tau_E}}.
\label{33}
\end{equation}
Still, $\tau_\alpha$ is of order $f_0 \tau_E$, what is physically
reasonable.

In this case $\zeta_E$ and $\rho_E$ vary with time as in
Eq.~(\ref{28}), whereas
\begin{equation}
\zeta_\alpha =\tau_\alpha ={\rm const}, \quad
\rho_\alpha=\frac{\rho_{0\alpha}}{X}. \label{34}
\end{equation}
The bulk viscosity corresponding to the $\alpha$- fluid component
is thus a constant. The density $\rho_\alpha$ decreases more
slowly with time than does $\rho_E \propto X^{-2}$.

\subsection{The case $\alpha=2$}

This case is quadratic modified gravity, in its simplest form.
From Eq.~(\ref{17}) we get the following cubic equation for $B$:
\begin{equation}
B^3+\left( 2+\frac{3}{4}\gamma \right)B^2+\frac{3}{2}\gamma
B-\frac{9}{8}T=0. \label{35}
\end{equation}
Let us put $\gamma=0$. If we draw a curve representing the
expression $(B^3+2B^2-9T/8)$ versus $B$, we see that it has a
local maximum at $B=-4/3$ and a local negative minimum at $B=0$.
There is thus one single positive root of the equation, for all
positive $T$. This root leads in turn to a viscosity-generated Big
Rip. A more detailed discussion of this case is given in Sect. 4.2
in \cite{brevik05b}. We only note here the time dependencies of
the $\alpha$- fluid component:
 \begin{equation}
\zeta_\alpha=\tau_\alpha \left(\frac{3H_0}{X}\right)^3, \quad
\rho_\alpha=\frac{\rho_{0\alpha}}{X^4}. \label{37}
\end{equation}

\section{Negative values of $\alpha:\, \alpha =-1$}

If $\alpha$ takes negative values, the situation becomes
qualitatively different in the following way: Positive values of
$B$ will always lead to a Big Rip singularity for the Hubble
parameter, as $H = H_0/X \rightarrow \infty$ at a finite value of
$t$. Similarly, both $\zeta_E \propto X^{-1}$ and $\rho_E \propto
X^{-2}$ will diverge. However, both $\zeta_\alpha \propto
X^{-(2\alpha-1)}$ and $\rho_\alpha \propto X^{-2\alpha}$ go to
{\it zero} at the Big Rip. The Einstein fluid component and the
$\alpha$-fluid component thus behave quite differently.

Let us consider $\alpha =-1$ as the most typical example:
\begin{equation}
S= \frac{1}{2\kappa^2}\int d^4 x \sqrt{-g}\left(
R+\frac{f_0}{R}+L_m \right). \label{38}
\end{equation}
From Eq.~(\ref{17}) we get the following equation for $B$:
\begin{equation}
B^2+\frac{1}{2}\frac{3-9\gamma +8T}{3+T}B-
\frac{1}{4}\frac{9\gamma-16T}{3+T}=0, \label{39}
\end{equation}
$T$ being being given by Eq.~(\ref{32}) as before.

Consider first the vacuum fluid case $\gamma=0$. The product
$B_1B_2$ of the two roots  is positive, whereas the sum
$(B_1+B_2)$ of them is negative. There is no positive root, thus
no Big Rip, in this case. The same is true for $\gamma <0$. The
case of general $\gamma$ is analyzed similarly by observing that
$B_1B_2$ has the same sign as $(16T-9\gamma)$, whereas $(B_1+B_2)$
has the same sign as $(9\gamma-3-8T)$. On physical grounds, we
expect that $T \ll 1$. It follows that for $\gamma <16T/9$ there
are two negative roots, whereas for $\gamma >16T/9$ there is one
positive and one negative root. In the latter case there is thus a
Big Rip possible, in the special sense explained above. Still,
there is no need of a scalar field.

\section{Summary. Remarks on Big Rip classification}

{\bf Summary.\,} The main purpose of this paper has been to
investigate whether the passage through the phantom barrier, from
the quintessence region ($w>-1$) into the phantom region ($w<-1$)
can be described as a viscosity-generated phenomenon in the
general case when the action is as in Eq.~(\ref{3}). As shown
earlier \cite{brevik05} such a mechanism works well in the case of
Einstein's gravity, if the bulk viscosity $\zeta \propto \theta$
where $\theta$ is the scalar expansion. Also, in the more general
case of pure modified gravity (Lagrangian of the form $R^\alpha$),
the same mechanism was found to work if $\zeta \propto
\theta^{2\alpha-1}$ \cite{brevik05a,brevik05b}, what is a natural
generalization of our basic ansatz.

The case where the action is as in Eq.~(\ref{3}) is more physical
than those considered in \cite{brevik05a,brevik05b} since the
Einstein component and the modified component are now combined
into one Lagrangian. It turns out that the system shows the same
kind of behavior as previously: there exists in principle a
viscosity-driven passage through the barrier $w=-1$. It becomes
however necessary to introduce a two-fluid model, since the bulk
viscosities for the Einstein component and the modified component
vary differently with time. Also, physical quantities such as the
density vary differently for the two fluid components.

Perhaps is this kind of behavior yet another indication of the
necessity of introducing a two-component fluid model in the late
universe, as emphasized, for instance, by Vikman within the
framework of the scalar field picture \cite{vikman05}.

It ought to be mentioned again that the the present model requires
the two coefficients $\tau_E$ and $\tau_\alpha$ to be related.
This requirement follows from Eqs.~(\ref{16}) and (\ref{17}).
There appears to be no direct physical reason why this should be
so; rather, the condition is a consequence of the mathematical
consistency of the formalism.

\bigskip
{\bf On the Big Rip classification.\,} There are actually several
variants of the Big Rip phenomenon, and it is of interest to trace
out to what category the singularities encountered in the present
paper belong. In Refs.~\cite{caldwell03,mcinnes02,barrow04} two
different types of Big Rip were discussed. In Ref.~\cite{nojiri05}
a classification of the known types of Big Rip was made and two
new types were discovered, so that there are according to this 
four types in all. Let $t=t_s$ be the instant when Big Rip occurs.
From Ref.~\cite{nojiri05} we reproduce the following
classification:

$\bullet$ Type I: For $t\rightarrow t_s,\, a\rightarrow \infty,\,
\rho \rightarrow \infty$ and $|p|\rightarrow \infty$

$\bullet$ Type II: For $t\rightarrow t_s, \, a\rightarrow
a_s,\,\rho \rightarrow \rho_s$ and $|p| \rightarrow \infty$

$\bullet$ Type III: For $t\rightarrow t_s, \, a\rightarrow a_s,\,
\rho \rightarrow \infty$ and $|p| \rightarrow \infty$

$\bullet$ Type IV: For $t\rightarrow t_s, \, a\rightarrow a_s,\,
\rho \rightarrow 0,\, |p|\rightarrow \infty$ and higher
derivatives of $H$ diverge.

Here $t_s, a_s$ and $\rho_s$ are constants with $a_s \neq 0.$ Type
I is the case one usually associates with the Big Rip concept,
emerging when $w<-1$.

In the present case, when $H$ is given as in Eq.~(\ref{12}), the
scale factor varies with time as
\begin{equation}
a(t)=\frac{1}{\left( 1-BH_0t \right)^{1/B}}. \label{40}
\end{equation}
Thus, if $B>0$, $a\rightarrow \infty$ when $t\rightarrow
t_s=1/(BH_0)$. It means that only Type I is an actual option in
our case. If the parameter $\alpha >0$ in the action (\ref{3}), we
have a Type I Big Rip both for the Einstein fluid component and
the $\alpha$-component. If $\alpha <0$, the Einstein component
still belongs to the Type I category, but the $\alpha$-component
belongs to a new type, different from the ones listed above, since
$\zeta_\alpha \rightarrow 0,\, \rho_\alpha \rightarrow 0, \,
p_\alpha =w \rho_\alpha  \rightarrow 0.$

\section*{Acknowledgments}

The present work originated from an idea put forward by Sergei D.
Odintsov. I thank him for his help. I acknowledge also valuable
information from various people responding to the previous papers
in this series \cite{brevik05,brevik05a,brevik05b}.

\end{document}